\documentclass[epsf,usegraphicx]{mn2e}

\usepackage{xspace}

\title[2XMMi J225036.9+573154 -- a new AM Her binary]
{2XMMi J225036.9+573154 -- a new eclipsing AM Her binary discovered
using XMM-Newton}

\author[]
{Gavin Ramsay$^{1}$, Simon Rosen$^{2}$, Pasi Hakala$^{3}$, 
Thomas Barclay$^{1,4}$\\
$^{1}$Armagh Observatory, College Hill, Armagh, BT61 9DG\\
$^{2}$Department of Physics and Astronomy, University of Leicester, 
University Road, Leicester LE1 7RH\\
$^{3}$Tuorla Observatory, University of Turku, V\"ais\"al\"antie 20, FIN-21500
Piikki\"o, Finland\\
$^{4}$Mullard Space Science Laboratory, University College London,
Holmbury St. Mary, Dorking, Surrey, RH5 6NT\\
}

\begin{document}
\outer\def\gtae {$\buildrel {\lower3pt\hbox{$>$}} \over 
{\lower2pt\hbox{$\sim$}} $}
\outer\def\ltae {$\buildrel {\lower3pt\hbox{$<$}} \over 
{\lower2pt\hbox{$\sim$}} $}
\newcommand{\ergscm} {erg s$^{-1}$ cm$^{-2}$}
\newcommand{\ergss} {erg s$^{-1}$}
\newcommand{\ergsd} {erg s$^{-1}$ $d^{2}_{100}$}
\newcommand{\pcmsq} {cm$^{-2}$}
\newcommand{\ros} {\sl ROSAT}
\newcommand{\chan} {\sl Chandra}
\newcommand{\xmm} {\sl XMM-Newton}
\def\rchi{{${\chi}_{\nu}^{2}$}}
\newcommand{\Msun} {$M_{\odot}$}
\newcommand{\Mwd} {$M_{wd}$}
\def\Mdot{\hbox{$\dot M$}}
\def\mdot{\hbox{$\dot m$}}
\newcommand{\teff}{\ensuremath{T_{\mathrm{eff}}}\xspace}
\newcommand{\src} {XMM J2250+5731}

\maketitle

\begin{abstract}
We report the discovery of an eclipsing polar, 2XMMi J225036.9+573154,
using {\xmm}. It was discovered by searching the light curves in the
2XMMi catalogue for objects showing X-ray variability. Its X-ray light
curve shows a total eclipse of the white dwarf by the secondary star
every 174 mins. An extended pre-eclipse absorption dip is observed in
soft X-rays at $\phi$=0.8--0.9, with evidence for a further dip
in the soft X-ray light curve at $\phi\sim$0.4. Further, X-rays are
seen from all orbital phases (apart from the eclipse) which makes it
unusual amongst eclipsing polars. We have identified the optical
counterpart, which is faint ($r$=21), and shows a deep eclipse ($>$3.5
mag in white light). Its X-ray spectrum does not show a distinct soft
X-ray component which is seen in many, but not all, polars. Its
optical spectrum shows H$\alpha$ in emission for a fraction of the
orbital period.
\end{abstract}


\section{Introduction}

Cataclysmic Variables are accreting binary systems in which a white
dwarf accretes material from a late type main sequence star through
Roche lobe overflow. If the white dwarf has a significant magnetic
field then the formation of an accretion disk can be disrupted or
prevented.  For white dwarfs with field strengths greater than
$\sim$10 MG, the accretion stream gets channelled onto the magnetic
poles where X-rays are emitted from the post-shock region. The
magnetic field also forces the spin period of the white dwarf to
synchronise with the binary orbital period. These accreting binaries
are called AM Her binaries or polars, since their optical emission is
strongly polarised.

The study of polars was transformed with the launch of the X-ray
satellite {\ros} in 1990.  Prior to this, around 17 systems were
known. {\ros} led directly to the discovery of around 30 new systems
(eg Beuermann \& Burwitz 1995).  It was expected that {\xmm}, launched
in 1999, would lead to the discovery of many more such
systems. Surprisingly, comparatively few have so far been discovered.

The 2XMM catalogue (Watson et al 2009) gives a description of
serendipitous X-ray sources discovered using the EPIC wide-field
instruments on board {\xmm}. This was followed by the release of the
2XMMi incremental catalogue which has 17 percent more discrete sources
than the 2XMM catalogue. Moreover, each source is accompanied by
source specific light curve and spectral products. In this paper we
report the discovery of an eclipsing polar, 2XMMi J225036.9+573154,
which was found as a result of searching the 2XMMi catalogue for
sources which showed variability in their X-ray light curve.

\section{{\xmm} observations}
\label{obs}

\subsection{The 2XMMi catalogue}

\begin{figure}
\begin{center}
\setlength{\unitlength}{1cm}
\begin{picture}(8,13.2)
\put(-0.5,-0.4){\includegraphics{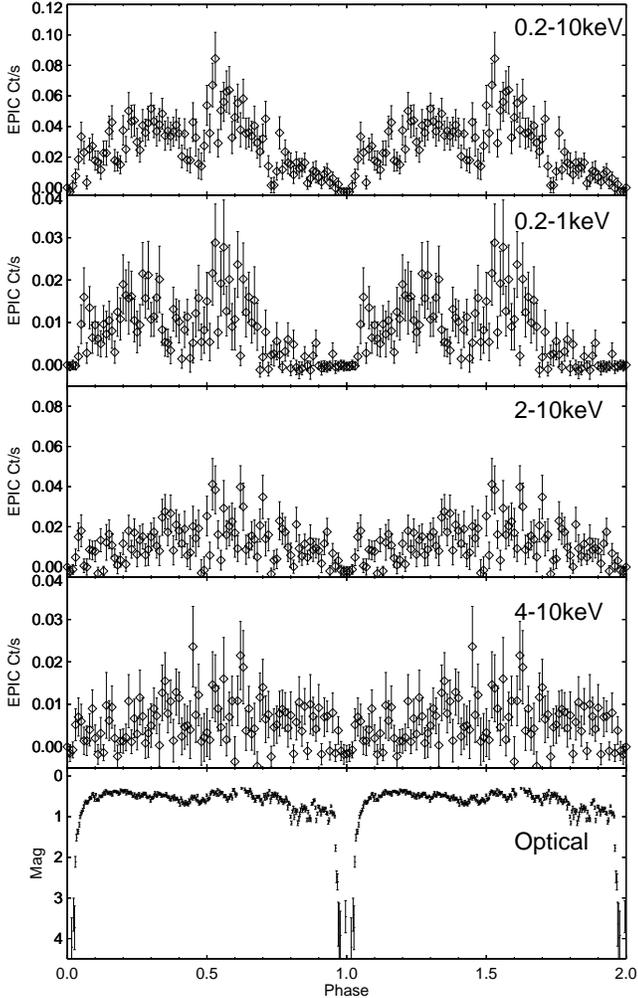}}
\end{picture}
\end{center}
\caption{The light curves of {\src} folded on a period of 174.2 mins
and $T_{o}$=2454124.245888 (TT). From the top we show the combined
EPIC pn plus EPIC MOS light curve in the 0.2--10keV energy band; the
0.2--1.0keV energy band; the 2--10keV energy band; the 4--10keV energy
band (binned into 100 bins) and in the lower panel the white light
data obtained using the NOT (the data have been folded but not
binned).}
\label{light}
\end{figure}

The 2XMMi catalogue has associated spectra and light curves that are
automatically extracted by the {\xmm} Science Survey Centre pipeline
processing software (Watson et al 2001) for sources with more than 500
counts in the EPIC detectors. An assessment of variability in the
individual light curves is made by determining {\rchi} of the data
about the mean, and then computing the consequent probability of the
constant (null) hypothesis. Those light curves for which this
probability is $< 10^{-5}$ are deemed variable. Sources which were
possibly compromised by further data quality issues were removed. An
initial search of the catalogue found around 400 sources which passed
these criteria. The light curves of these sources were visually
inspected for periodic behaviour. One source, 2XMMi J225036.9+573154
(hereafter XMM J2250+5731), was found which showed a characteristic
repeating shape on a period of $\sim$174 min (Figure \ref{light}).

\subsection{Pointed observations}

{\src} was found in the field of G107.5-1.5 which was observed on 23rd
Jan 2007. The EPIC detectors were each configured in full window mode
and used the medium filter. The field was observed for a total of 32.9
ksec in the EPIC pn detector and 34.5 ksec in both EPIC MOS
detectors. The source was just outside the field of view of the
Optical Monitor.  Since the source was towards the edge of the EPIC
detectors, and there was a nearby (28$^{''}$) X-ray source, XMM
J225037.9+573127, which appears to be an active late-type star, we
extracted the data from the {\xmm} archive and re-extracted the X-ray
light curves and spectra of {\src}.

The data were processed using {\xmm} SAS v8.0.1 (released Oct 2008).
Only X-ray events which were graded as {\tt PATTERN}=0-4 and {\tt
FLAG}=0 were used. Events were extracted from a circular aperture with
10$^{''}$ radius centred on the source, with background events being
extracted from source free areas on the same chip as the source. The
background data were scaled to give the same area as the source
extraction area and subtracted from the source area. (We estimate that
the nearby source XMM J225037.9+573127 contributes around 1.5 percent
of the flux below 2keV, and a negligible amount at energies above
4keV). To ensure that the spectra were correctly flux calibrated we
produced detector spectral response files and ancillary files using
the SAS tasks {\tt rmfgen} and {\tt arfgen} respectively.

\section{The X-ray light curves}
\label{sec-light}

We extracted light curves of {\src} in the 0.2--10keV, 0.2--1.0keV,
2--10keV and 4--10keV energy bands from the EPIC pn, EPIC MOS1 and
EPIC MOS2 detectors using the method described above. We then obtained
a combined light curve for each energy band by adding the separate
light curves. Each light curve shows a distinctive sharp drop in
intensity every 174 min. This is due to the secondary star eclipsing
the accretion region(s) on the white dwarf and represents the binary
orbital period. The observation covers 3 eclipses.

\begin{figure}
\begin{center}
\setlength{\unitlength}{1cm}
\begin{picture}(8,5.5)
\put(-0.5,-0.2){\includegraphics{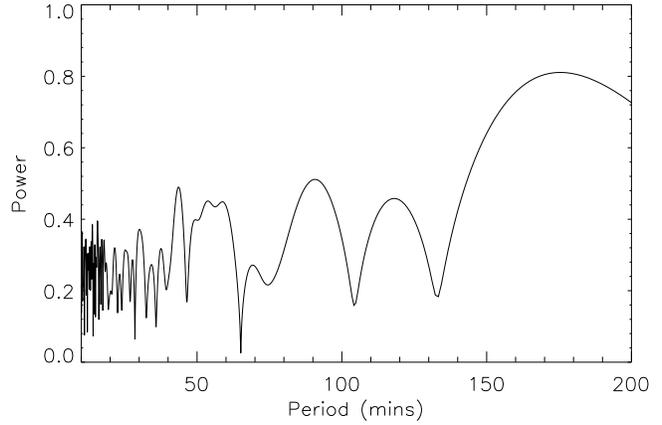}}
\end{picture}
\end{center}
\caption{The power spectrum of the combined EPIC (pn plus MOS)
0.2--10keV light curve.}
\label{power}
\end{figure}

We used the standard Lomb-Scargle power spectrum analysis to search
for periods in the data (Figure \ref{power}). The error on the period
was then determined using a bootstrap approach incorporating the
generation of synthetic light curves. We find that the period is
0.1210$\pm$0.0018 days (=174.2$\pm$2.6 mins). We folded the light
curve in each of the 4 energy bands on this period and show these
light curves in Figure \ref{light}.

We have phased the data so that the eclipse, which is total in each
energy band, defines $\phi$=0.0. {\src} is relatively faint in
X-rays, reaching a peak of $\sim$0.08 ct/s in the combined EPIC
0.2--10keV light curve, although this count rate has not been
corrected for the source being far off-axis.

Prior to the eclipse, there is a marked decrease in soft X-ray photons
over the phase range $\phi\sim$0.7--1.0, compared to those at higher
energies. This phenomenon has been seen in other polars (eg Watson et
al 1989) and occurs when the accretion stream obscures our view of the
hot accretion region located in the upper hemisphere on the white
dwarf. Compared to V2301 Oph (for instance, Ramsay \& Cropper 2007),
the `pre-eclipse' dip seen in {\src} is more extended suggesting that
material gets lifted out of the orbital plane over a wider range in
azimuth.

At softer energies ($<$1keV) there is also a dip in the light curve
centered at $\phi\sim$0.4 and with a duration of $\sim$0.1--0.2
cycles. At higher energies, there is no obvious broad dip at these
orbital phases although there are a couple of bins between
$\phi$=0.4--0.5 which are consistent with zero counts. However, since
other bins with negligible flux are also seen at different phases this
may just be due to low counting statistics. This dip could either be
due to a second dip caused by an accretion stream or it could be due
to the rotation of the accretion regions rotating into and out of view
as the white dwarf rotates. We will discuss this further in \S
\ref{discussion}.

\section{X-ray spectral fits}

We extracted spectra from each EPIC detector in the manner described
in \S \ref{obs}. Initially we extracted spectra using all the
available data.  However, since the light curves (cf Figure
\ref{light}) imply the presence of a pre-eclipse dip, we then
extracted spectra from the phase interval which was not strongly
affected by absorption, ie $\phi\sim$0.05--0.7. We also exclude the
phase interval $\phi$=0.38--0.5 which could also be affected by
absorption (\S 3).

In polars, X-rays are generated in a post-shock region at some height
above the photosphere of the white dwarf.  Since the X-ray spectrum of
{\src} has a relatively low signal to noise compared to many polars
previously studied using {\xmm} (eg Ramsay \& Cropper 2004), we used a
simple single temperature thermal bremsstrahlung emission model rather
than a more complex (and more physical) stratified cooling flow model
(eg Cropper et al 1998, 1999).

\begin{figure}
\begin{center}
\setlength{\unitlength}{1cm}
\begin{picture}(8,6)
\put(-0.5,-0.8){\includegraphics{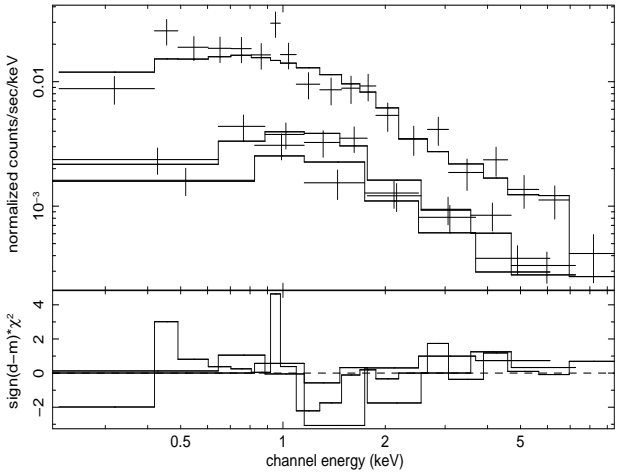}}
\end{picture}
\end{center}
\caption{Upper Panel: The EPIC spectra extracted from the bright phase
interval along with the best fits (the upper most spectrum is that
derived from the EPIC pn, while the lower ones are derived from the
EPIC MOS 1 and 2 detectors). Lower Panel: the residuals to the best
fit are shown in units of $\chi^{2}$.}
\label{xray-spec}
\end{figure}

We used the {\tt XSPEC} package (Arnaud 1996) to fit the X-ray
spectra.  We fitted all three EPIC spectra simultaneously and tied the
spectral parameters apart from the normalisation parameters. We used
the {\tt tbabs} absorption model (the T\"{u}bingen--Boulder absorption
ISM model, Wilms, Allen \& McCray 2000), a single temperature thermal
bremsstrahlung component with temperature fixed at $kT=20$keV. We
added a Gaussian component to account for any emission between
6.4--6.8keV.  The spectra along with the best fit (\rchi=1.12) are
shown in Figure \ref{xray-spec}.  We show the spectral parameters, the
observed and unabsorbed bolometric fluxes in Table \ref{fits}. Due to
the low signal to noise of the spectra, the equivalent width of the Fe
K$\alpha$ emission line features was poorly constrained.

In many polars, a strong soft X-ray component ($kT_{bb}\sim$40eV) is
seen in their X-ray spectra (eg Ramsay et al 1994, Beuermann \&
Burwitz 1995).  This is due to the hard X-rays irradiating the
photosphere of the white dwarf which are then re-emitted as soft
X-rays. The standard accretion model predicts that
$L_{soft}/L_{hard}\sim$0.5, where $L_{soft}$ and $L_{hard}$ are the
luminosities of the soft and hard X-ray components respectively (Lamb
\& Masters 1979, King \& Lasota 1979). Although the energy balance in
polars was a source of great debate for many years, Ramsay \& Cropper
(2004) showed that the majority of polars in a high accretion state
have X-ray spectra which are in good agreement with the standard
model.

To see if such a soft X-ray component could be `hidden' by the
moderate level of absorption (cf Table \ref{fits}) we added a
blackbody with a range of different temperatures. We fixed its
normalisation so that the implied ratio,
$L_{soft}/L_{hard}\sim$0.5. Since the soft X-rays are optically thick,
and hence the intrinsic soft X-ray luminosity is viewing angle
dependant, we assumed a viewing angle of 45$^{\circ}$ for argument. If
we just consider the X-ray data, we find a blackbody with temperature
less than $kT$\ltae 20eV can easily be hidden.

We were fortunate in being able to obtain a short observation of the
field of {\src} using {\sl Swift} on the 3rd and 4th Dec
2008. Observations using the UV Optical Telescope (Roming et al 2005)
were made using the UVW2 filter (peak effective wavelength
2120\AA). {\src} was not detected, and we estimated a 3$\sigma$ upper
limit of $\sim2.4\times10^{-17}$ \ergscm \AA$^{-1}$. If we assume a
blackbody of different temperatures and with a normalisation such that
$L_{soft}/L_{hard}\sim$0.5, we find that a blackbody of
$kT\sim$5--20eV can be present and not detected in the near UV or soft
X-ray energy ranges. (Although a handful of X-ray events were detected
near the source position of {\src}, they were too low to derive any
meaningful information).

The unabsorbed bolometric flux implies an X-ray luminosity of
$\sim8\times10^{29} d_{100}^{2}$ erg/s, where $d_{100}^{2}$ is the
distance in units of 100 pc. Ramsay \& Cropper (2004) found that the
mean bolometric luminosity in their sample of polars observed in a
high state using {\xmm} was $\sim2\times10^{32}$ erg/s. In the
next section we find that {\src} shows a range in optical brightness
over the longer term and therefore a range of accretion states (a
general characteristic of polars). Assuming that {\src} was observed
in a high accretion state at the epoch of the {\xmm} observations we
find that in order that {\src} has an X-ray luminosity consistent with
other polars in a high state it must lie at a distance of
$\sim$1.5--2.0 kpc. With Galactic co-ordinates of $l=107.2^{\circ}$
and $b=-1.6^{\circ}$, this places {\src} close to the Perseus spiral
arm (Xu et al 2005).

\begin{table}
\begin{center}
\begin{tabular}{lr}
\hline
$N_{H}$ & $3.4^{+2.2}_{-1.8}\times10^{20}$ \pcmsq \\
Flux$^{o}$ EPIC pn & $3.5^{+0.3}_{-0.4}\times10^{-13}$ \ergscm \\
Flux$^{o}$ EPIC mos1 & $2.8^{+0.4}_{-0.4}\times10^{-13}$ \ergscm \\
Flux$^{o}$ EPIC mos2 & $2.1^{+0.4}_{-0.3}\times10^{-13}$ \ergscm \\
Flux$^{u}$ EPIC pn & $8.1^{+0.9}_{-1.0}\times10^{-13}$ \ergscm \\
Flux$^{u}$ EPIC mos1 & $6.4^{+1.0}_{-0.9}\times10^{-13}$ \ergscm \\
Flux$^{u}$ EPIC mos2 & $4.8^{+0.9}_{-0.8}\times10^{-13}$ \ergscm \\
\rchi & 1.12 (34 dof)\\
\hline
\end{tabular}
\end{center}
\caption{The spectral fit to the EPIC pn, MOS1 and MOS2 spectra
extracted from $\phi$=0.05--0.7. Flux$^{o}$ refers to the observed
flux measured over the 0.2--10keV energy band and Flux$^{u}$ refers to
the unabsorbed bolometric flux.}
\label{fits}
\end{table}

\section{Optical photometry}

\begin{figure}
\begin{center}
\setlength{\unitlength}{1cm}
\begin{picture}(8,8)
\put(0,0){\includegraphics{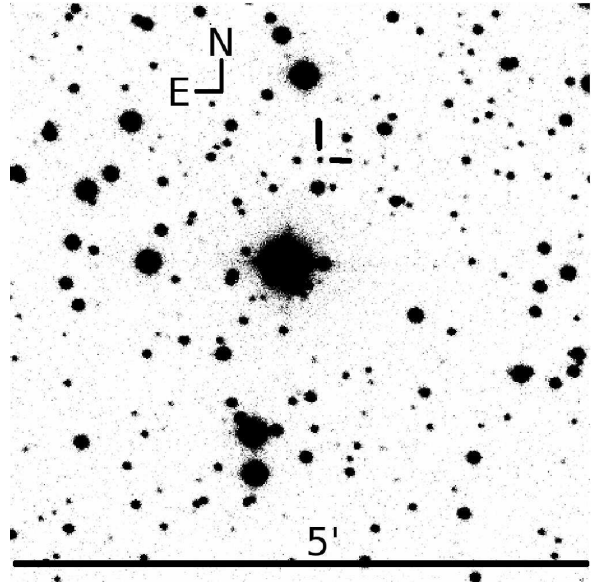}}
\end{picture}
\end{center}
\caption{The finding chart of {\src} extracted from a $g$ band image
taken with the INT WFC on 6th Nov 2008.}
\label{finding}
\end{figure}

To locate the optical counterpart of {\src} we obtained optical
photometry using ALFOSC on the Nordic Optical Telescope (NOT) located
on La Palma during 28th Sept 2008. Each exposure was in `white light'
and 15 sec in length, with another 5 sec of readout time, resulting in
2.9 h of data in total. Each source in the field (Figure
\ref{finding}) was searched for variability. One source showed a clear
eclipse lasting for $\sim$12 mins and a depth \gtae 3.5 mag (Figure
\ref{light}). This is the optical counterpart to {\src} and its
co-ordinates are $\alpha_{2000}=22^{h} 50^{m} 36.97^{s},
\delta_{2000}=+57^{\circ} 31^{'} 54.2^{''}$ (which is within
0.8$^{''}$ of the X-ray position).

We searched the IPHAS catalogue (Drew et al 2005) which surveyed the
northern galactic plane in $r,i,$H$\alpha$ filters to determine if the
optical counterpart of {\src} was detected in this survey.  We find
that IPHAS gives $r=20.32\pm0.05, i=20.1\pm0.2, H\alpha=19.69\pm0.09$
for {\src}.  All sources within a 30$^{''}$ radius of the X-ray
position were extracted. {\src} is at the extreme blue end in the
$(r-i)$ distribution and consistent with the location of the CVs found
in IPHAS data in the $(r-i), (r-H\alpha)$ colour-colour plane (cf fig1
of Corradi et al 2008).

We also obtained $U, g, R$ images of {\src} using the Wide Field
Camera on the Isaac Newton Telescope on 6th Nov 2008: Figure
\ref{finding} shows the $g$ band image of the immediate field. Using
standard star observations taken immediately before these observations
we find that $U=21.75\pm$0.11, $g=21.16\pm0.05$ and
$r=21.53\pm$0.05. Compared to other stars in the field, it is clearly
blue and appears to be more than 1 mag fainter than found at the epoch
of the IPHAS pointings. This is not unexpected since polars are known
to show different accretion states.

\section{Optical spectroscopy}

We obtained spectra of {\src} using the 4.2m William Herschel
Telescope and the Intermediate dispersion Spectrograph and Imaging
System (ISIS) on La Palma on 6th Oct 2008. We used the R300B and R158R
gratings giving a spectral resolution of $\sim$2.5\AA \hspace{1mm} and
$\sim$5\AA \hspace{1mm} respectively. The seeing was $\sim0.8^{''}$
and the slit was set to match the seeing. We took 16 spectra in both
the red and blue arms.

\begin{figure}
\begin{center}
\setlength{\unitlength}{1cm}
\begin{picture}(8,5.5)
\put(-0.5,-0.2){\includegraphics{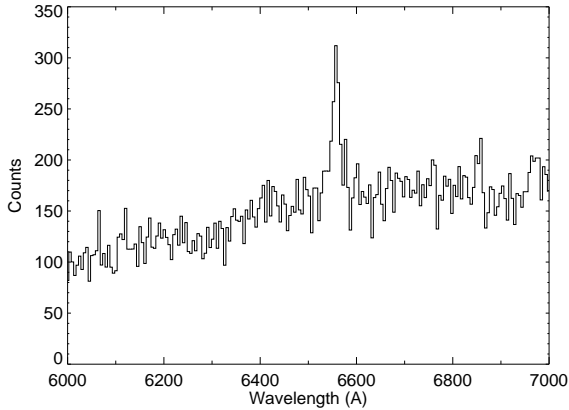}}
\end{picture}
\end{center}
\caption{In 9 our of our 16 spectra taken of {\src} using the WHT
and ISIS in Oct 2008, we detected the H$\alpha$ emission
line. This figure shows the mean of these 9 spectra. ISIS and the WHT 
on 6th Oct 2008. The flux scale is arbitrary.}
\label{red-spec}
\end{figure}

With an out of eclipse brightness of $r\sim$21, each individual
spectrum was of low signal to noise. Moreover, in the blue arm, there
was electronic noise in the images, the pattern of which varied from
image to image. This coupled with the low signal to noise of the
spectra prevented us from extracting any useful information from the
blue arm. In the red arm, we were able to extract a spectrum from each
image. For 9 sequential spectra we were able to detect H$\alpha$ in
emission. We show the mean of these spectra in Figure
\ref{red-spec}. For the remaining 7 spectra for which we did not
detect H$\alpha$ in emission we attribute this to the fact that the
observations occurred during the phase interval of the pre-eclipse
absorption dip or that the accretion stream was presenting a small
surface area at those phase intervals. (Given the error on the
orbital period, \S 3, the phasing of the WHT spectra using the NOT
photometric observations as a marker of the phasing is uncertain by
approximately one orbital cycle).

\section{Discussion}
\label{discussion}

\subsection{The X-ray light curve}

In polars, it is thought that the magnetic axis of the white dwarf is
tilted towards the secondary star, but shifted a few 10's of degrees
ahead in azimuth (as the binary rotates) of the line of center joining
the two stars (eg Cropper 1988). It is therefore the accretion region
in the upper hemisphere which is obscured by the accretion flow during
the pre-eclipse absorption dip.

Eclipsing polars have long been the target of dedicated X-ray
observations. Many of these polars show a distinct bright and faint
phase as the accretion region rotates into view and out of view and
many show a characteristic pre-eclipse absorption dip. In these
systems there is no evidence for a second accretion pole. One of the
few eclipsing polars to show emission throughout the binary phase is
V2301 Oph (Ramsay \& Cropper 2007). We find that in the case of
{\src}, X-ray emission is also seen throughout the orbital phase.
We attempted to invert the X-ray light curves and map the X-ray
regions on the white dwarf using an approach similar to that of
Cropper \& Horne (1994). However, because of the relatively low signal
to noise of the data we could not identify a unique solution.

In \S 3 we noted the presence of a broad dip in soft X-rays at
$\phi\sim$0.4 which could be attributed to either an accretion stream
(since there is no similar feature at higher energies) or the rotation
of the accretion region(s) as they come into and out of view. In the
former case, the dip could be due to a second accretion stream
obscuring our line of sight to the accretion region located in the
lower hemisphere of the white dwarf. To our knowledge this would make
{\src} unique amongst polars in showing two absorption dips. In the
latter case, the change in the soft X-ray light curve could be due to
either the rotation of two accretion regions, located in opposite
hemispheres, or the rotation of one relatively large polar
region. (Our inversion maps showed that both scenarios could
re-produce the soft X-ray light curves). The fact that soft X-rays
emitted at the base of the accretion region are optically thick and
hence viewing angle dependant could account for the change in the soft
X-ray flux. In contrast, the harder X-rays are optically thin and
therefore not viewing angle dependant.

Optical polarimetry data would be able to confirm the presence of two
accretion poles. However, since {\src} is rather faint, this may prove
challenging.  

\subsection{The energy balance}

Ramsay \& Cropper (2004) presented the results of a snap-shot survey
of polars observed in a high accretion state using {\xmm}. They found
that 7 out of 21 systems did not show a distinct soft X-ray component.
Vogel et al (2008) also report that 2XMMp J131223.4+173659, which was
discovered serendipitously using {\xmm}, does not show a soft X-ray
component. We have searched the literature for further observations of
polars observed using {\xmm} in a high state: we find an additional 6
polars. (We are aware of a number of observations of polars in the
high state which have been carried out but have not as of yet been
published).  V1309 Ori (Schwarz et al 2005), V1432 Aql (Rana et al
2005) and SDSS J075240.45+362823.2 (Homer et al 2005) all show
distinct soft X-ray components while SDSS J072910.68+365838.3 and SDSS
J170053.30+400357.6 (Homer et al 2005) do not. In the case of SDSS
J015543.4+002807.2 (Schmidt et al 2005) the existence of a soft
component is not required at a high significance and hence we define
it as not having a soft X-ray component. We therefore find that 10 out
of 27 systems observed in a high state do not show a distinct soft
X-ray component.

Ramsay \& Cropper (2007) suggested that if the temperature of the
re-processed X-rays was low enough, it would not be observable using
the {\xmm} X-ray detectors. This view is also supported by the
analysis carried out by Vogel et al (2008) on observations of 2XMMp
J131223.4+173659.  The reason for this could be that the accretion
flow covers a larger fraction of the photosphere of the white dwarf or
that the mass accretion rate is lower than in systems which showed a
soft component (since $kT_{bb}\propto({\dot{M}/f}^{1/4})$, where
$\dot{M}$ is the mass accretion rate and $f$ is the fractional area
over which accretion is occurring).

There is no obvious reason as to why some polars would have accretion
occurring over a larger area than others: they share no common
characteristics such as magnetic field strength or orbital
period. Indeed, as noted by Ramsay \& Cropper (2004) two systems (BY
Cam and RX J2115--58) have one pole which shows a soft component and
one pole which does not. Further, three systems which have at least
one pole which does not show a soft component are asynchronous systems. 
However, V1432 Aql which does show a soft
component is also an asynchronous polar.

\section{Conclusions}

We have serendipitously discovered a faint polar, {\src}, with an
orbital period of 2.9 h, in the 2XMMi catalogue. We have identified
the optical counterpart as a $r\sim21$ object and it shows a deep
eclipse in the optical and X-ray bands lasting $\sim$12 mins. At soft
X-ray energies there is a distinctive drop in counts starting
$\sim$0.3 cycles before the eclipse. This is due to the accretion
stream obscuring the accretion region in the upper hemisphere of the
white dwarf. A second dip is seen in soft X-rays at $\phi\sim$0.4
which could either be due to obscuration of the accretion region by a
second stream or due to the rotation of the accretion region(s)
rotating into and out of view. Amongst eclipsing polars, {\src} is
unusual in that X-ray emission is visible over the whole of the binary
orbital phase, apart from the eclipse.

We have analysed the X-ray spectrum of {\src} and find no evidence for
a distinct soft X-ray component. Of the 27 polars which have been
observed using {\xmm} and found to be in a high accretion state, 10
show no distinct soft X-ray component. This is a surprisingly high
fraction.  This together with the result that only a small fraction of
polars show a soft X-ray excess (Ramsay \& Cropper 2004), changes our
whole perception of polars being strong soft X-ray sources. Further,
it suggests that polars with strong soft X-ray components were
preferentially discovered using {\sl EXOSAT}.

\section{Acknowledgements}

Based on observations obtained with {\xmm}, an ESA science mission
with instruments and contributions directly funded by ESA Member
States and NASA. We thank Gillian James for providing an initial
reduction of the {\xmm} data and Diana Hannikainen and Hanna Tokola
for assisting with the NOT observations. Observations were made using
the William Herschel Telescope, the Isaac Newton Telescope and the
Nordic Optical Telescope on La Palma. We gratefully acknowledge the
support of each of the observatories staff. We also thank Andrew
Beardmore and other members of the {\sl Swift} team for scheduling
observations of our target. Some of the data presented here have been
taken using ALFOSC, which is owned by the Instituto de Astrofisica de
Andalucia (IAA) and operated at the Nordic Optical Telescope under
agreement between IAA and the NBIfAFG of the Astronomical Observatory
of Copenhagen.


\begin{thebibliography}{}
\bibitem{}Arnaud, K. A., 1996, Astronomical Data Analysis Software and 
Systems V, eds Jacoby, G., Barnes, J., p17, ASP Conf Series, 101
\bibitem{}Beuermann, K., Burwitz, V., 1995, In ASP Conf Series, 85, 99
\bibitem{}Corradi, R. L. M., et al, 2008, A\&A, 480, 409
\bibitem{}Cropper, M., 1988, MNRAS, 231, 597
\bibitem{}Cropper, M., Horne, K., 1994, MNRAS, 267, 481
\bibitem{}Cropper, M., Ramsay, G., Wu, K.,  1998, MNRAS, 293, 222
\bibitem{}Cropper, M., Wu, K., Ramsay, G., Kocabiyik, A.,1999, MNRAS, 306, 684
\bibitem{}Drew, J., et al,  2005, MNRAS, 362, 753
\bibitem{}Homer, L., et al, 2005, ApJ, 620, 929
\bibitem{}King A. R., Lasota J. P., 1979, MNRAS, 188, 653
\bibitem{}Lamb D. Q., Masters A. R., 1979, ApJ, 234, 117
\bibitem{}Ramsay, G., Mason, K. O., Cropper, M., Watson, M. G., Clayton, 
K. L., 1994, MNRAS, 270
\bibitem{}Ramsay, G., Cropper, M., 2004, MNRAS, 347, 497
\bibitem{}Ramsay, G., Cropper, M., 2007, MNRAS, 379, 1207
\bibitem{}Rana, V. R., Singh, K. P., Barrett, P. E., Buckley, D. A. H., 
2005, ApJ, 625, 351
\bibitem{}Roming, P. W. A., et al, 2005, Space Sci Rev, 120, 95
\bibitem{}Schmidt, G. D., et al, 2005, ApJ, 620, 422
\bibitem{}Schwarz, R., Reinsch, K., Beuermann, K., Burwitz, V., 2005,
A\&A, 442, 271
\bibitem{}Vogel, J., Byckling, K., Schwope, A., Osborne, J. P., 
Schwarz, R., Watson, M. G., 2008, A\&, 485, 787
\bibitem{}Watson, M. G., King, A. R., Jones, M. H., Motch, C., 1989, MNRAS,
237, 299
\bibitem{}Watson, M. G., et al, 2001, A\&A, 365, L51
\bibitem{}Watson, M. G., et al, 2009, A\&A, 493, 339 
\bibitem{}Wilms, J., Allen, A., McCray, R., 2000, ApJ, 542, 914
\bibitem{}Xu, Y., Reid, M. J., Zheng, X. W., Menten, K. M., 2006, 311, 54
\end{thebibliography}
\end{document}